\documentclass{article}
\usepackage{arxiv}
\usepackage{multirow} 
\usepackage{tabularx}
\usepackage{makecell}
\usepackage[utf8]{inputenc} 
\usepackage[T1]{fontenc}    
\usepackage[portuguese]{babel}  
\usepackage{hyperref}       
\usepackage{url}            
\usepackage{booktabs}       
\usepackage{amsfonts}       
\usepackage{nicefrac}       
\usepackage{microtype}      
\usepackage{graphicx}
\usepackage{natbib}
\usepackage{doi}
\usepackage[xcolor=dvipsnames,
todonotes={colorinlistoftodos,
                prependcaption,
                textsize=small,
                backgroundcolor=orange!10,
                textcolor=black,
                linecolor=orange,
                bordercolor=orange}]{changes}

\title{Socio-spatial segregation and human mobility: A review of empirical evidence}

\author{ \href{https://orcid.org/0000-0002-6982-1654}{\includegraphics[scale=0.06]{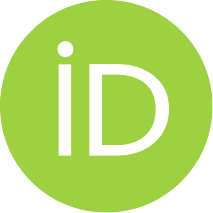}\hspace{1mm}Yuan~Liao}\thanks{Corresponding author. Also affiliated with Department of Applied Mathematics and Computer Science, Technical University of Denmark, Lyngby, Denmark.} \\
	Department of Space, Earth and Environment\\
	Chalmers University of Technology\\
	Gothenburg, Sweden \\
	\texttt{yuan.liao@chalmers.se} \\
	\And
	\href{https://orcid.org/0000-0001-6671-2578}{\includegraphics[scale=0.06]{orcid.pdf}\hspace{1mm}Jorge~Gil} \\
	Department of Architecture and Civil Engineering\\
	Chalmers University of Technology\\
	Gothenburg, Sweden\\
	\texttt{jorge.gil@chalmers.se} \\
    \And
	\href{https://orcid.org/0000-0002-4852-1177}{\includegraphics[scale=0.06]{orcid.pdf}\hspace{1mm}Sonia~Yeh} \\
	Department of Space, Earth and Environment\\
	Chalmers University of Technology\\
	Gothenburg, Sweden\\
	\texttt{sonia.yeh@chalmers.se} \\
    \And
	\href{https://orcid.org/0000-0003-2125-7465}{\includegraphics[scale=0.06]{orcid.pdf}\hspace{1mm}Rafael~H. M.~Pereira} \\
	Institute for Applied Economic Research (Ipea) - Brazil\\
	Data Science Division, Brazil\\
	\texttt{rafael.pereira@ipea.gov.br} \\
    \And
	\href{https://orcid.org/0000-0001-6003-1165}{\includegraphics[scale=0.06]{orcid.pdf}\hspace{1mm}Laura~Alessandretti} \\
	Department of Applied Mathematics and Computer Science\\
	Technical University of Denmark\\
    Lyngby, Denmark\\
	\texttt{lauale@dtu.dk} \\ 
}

\renewcommand{\shorttitle}{Socio-spatial segregation and human mobility: A review of empirical evidence}

\hypersetup{
pdftitle={A template for the arxiv style},
pdfsubject={q-bio.NC, q-bio.QM},
pdfauthor={David S.~Hippocampus, Elias D.~Striatum},
pdfkeywords={First keyword, Second keyword, More},
}

\begin{document}
\newcommand{\tabincell}[2]{
\begin{tabular}{@{}#1@{}}#2\end{tabular}
}
\maketitle

\begin{abstract}
Socio-spatial segregation is the physical separation of different social, economic, or demographic groups within a geographic space, often resulting in unequal access to resources, services, and opportunities.
The literature has traditionally focused on residential segregation, examining how individuals' residential locations are distributed differently across neighborhoods based on various social attributes, e.g., race, ethnicity, and income.
However, this approach overlooks the complexity of spatial segregation in people's daily activities, which often extend far beyond residential areas.
Since the 2010s, emerging mobility data sources have enabled a new understanding of socio-spatial segregation by considering daily activities such as work, school, shopping, and leisure visits.
From traditional surveys to GPS trajectories, diverse data sources reveal that daily mobility can result in spatial segregation levels that differ from those observed in residential segregation.
This literature review focuses on three critical questions: (a) What are the strengths and limitations of segregation research incorporating extensive mobility data? (b) How do human mobility patterns relate to individuals' residential vs. experienced segregation levels? and (c) What key factors explain the relationship between one's mobility patterns and experienced segregation?
Our literature review enhances the understanding of socio-spatial segregation at the individual level and clarifies core concepts and methodological challenges in the field. 
Our review explores studies of key themes: segregation, activity space, co-presence, and the built environment.
By synthesizing their findings, we aim to offer actionable insights for reducing segregation.
\end{abstract}


\section{Introduction}
Socio-spatial segregation is the physical separation of different social, economic, or demographic groups within a geographic space, often resulting in unequal access to resources, services, and opportunities. 
Socio-spatial segregation manifests as a distinct, uneven distribution of these groups across different geographical areas and is often characterized by limited social interactions.
Understanding segregation holds significant importance in our increasingly urbanized planet. 
Sustainable urban development fosters diverse populations and promotes social cohesion by facilitating access to vital resources, public services \citep{joelsson2022cracks}, educational \citep{zhang2022residential}, and employment opportunities \citep{silm2014ethnic} to all population groups.  
However, segregated cities can lead to differentiated access to such services and opportunities, perpetuating disparities in economic, social, and health outcomes \citep{li2022towards, hu2022examining,xu2023segregation}. 
Furthermore, high levels of segregation mean fewer opportunities for individuals from different backgrounds to come into contact with each other, resulting in limited opportunities for group interaction and exposure \citep{moro2021mobility}. \par

Quantitative studies on socio-spatial segregation focus on the geographic separation or clustering of social groups within physical spaces \citep{li2022towards}.
They assume the precursor of social interaction is being in the same place, i.e., the co-presence of individuals \citep{rokem2018segregation} or framed as socioeconomic mixing and exposure among diverse individuals \citep{nilforoshan2023human}. 
Socio-spatial segregation is a long-standing research topic deeply rooted in urbanization history, significantly advanced by the Chicago School of Sociology \citep[e.g.,][]{park2019city}.
These scholars have measured segregation from a static standpoint by looking at how residential locations are sorted into different neighborhoods based on income, race, and education etc \citep{feitosa2007global}.
The residential environment is a solid basis for studying socio-spatial segregation, as it strongly influences individuals' access to key urban resources.
However, a large body of recent work shows that residential segregation alone cannot fully capture the complexity of spatial segregation in urban areas \citep{netto2001socio}. 
Thus, understanding socio-spatial segregation requires considering other activity locations beyond residential spaces \citep{kwan2013beyond, silm2014ethnic}. 
These include locations visited for work \citep{zhou2021workplace}, school, shopping, and leisure \citep{toger2023inequality} both indoor and outdoor, as well as locations visited when en route, i.e., in the process of reaching these destinations. \par

To engage in out-of-home activities, people need to move outside where they live, making mobility a key aspect of understanding socio-spatial segregation.
Over the past one to two decades, many studies have advanced a dynamic understanding of segregation by incorporating activity spaces \citep{sampson2020beyond, candipan2021residence}, considering the geography of individuals' daily activity and mobility patterns beyond their residential area, thus providing a more comprehensive understanding of socio-spatial segregation. 
This enhanced understanding largely stems from the widespread availability of extensive human mobility data, shedding light on how individuals allocate their time among various activity locations.
The activity space approaches are driven by empirical mobility data from traditional data sources, such as travel surveys \citep{li2017measuring}, and emerging sources of big geolocation data covering large populations and geographical extent, e.g., geolocation tracking devices \citep{roulston2013gps}, mobile phone data \citep{silm2014ethnic,xu2019quantifying}, or social media platforms \citep{wang2018urban,candipan2021residence}. \par

Despite the growing number of studies considering activities beyond residential areas, there is a lack of a comprehensive review of the empirical findings on how individuals' segregation levels measured based on activity space, i.e., experienced segregation \citep{moro2021mobility,athey2021estimating,wu2023revealing,xu2024experienced} relate to residential segregation.
The literature has shown contradictory findings: socio-spatial segregation considering activity locations outside the residence can be lower or higher than individuals' residential segregation level \citep{kwan2013beyond}. 
The answer to this question lies in mobility, i.e., the movement of people from place to place via the built environment, e.g., transport modes, which connect people between their residences to activity locations outside the home.
In exploring the role of individual mobility in socio-spatial segregation patterns based on the existing literature, we attempt to answer three questions:
\begin{itemize}
\item What are the strengths and limitations of segregation research incorporating extensive mobility data?
\item How do human mobility patterns relate to individuals' segregation levels, i.e., residential vs. other activity places?
\item What key factors explain the relationship between one's mobility patterns and experienced segregation?
\end{itemize}
Answering these questions can contribute to providing actionable insights for reducing segregation and social inequalities. \par

In this paper, we review socio-spatial segregation research through the prism of individual mobility, drawing empirical evidence from the themes of segregation, activity space, co-presence, and the built environment (Section \ref{sec:methodology}).
Based on the literature, we first define socio-spatial segregation and its quantification (Section \ref{sec:measure}), followed by a critical reflection on methodologies in existing studies (Section \ref{sec:methods}).
We then review studies based on activity space approaches, including evidence on the relationship between mobility and experienced segregation levels (Section \ref{sec:exp}).
In Section \ref{sec:fact_prom}, we further draw findings from the built environment research to discuss critical factors explaining experienced segregation differing between population groups.
Finally, we synthesize these findings, highlight research gaps, and suggest directions for future research (Section \ref{sec:discussion}). \par

\section{Methods}\label{sec:methodology}
This literature review is centered around three pivotal concepts: spatial segregation, activity space, and the role of the built environment in facilitating mobility.
We design a list of keywords and search queries around four themes: segregation, activity space, co-presence, and built environment (Table \ref{tab:keywords}). 
The included studies were extracted from the Scopus database on Oct 18, 2023, and processed to answer the three research questions.
They cover 176 original articles in English published in journals or conferences.
These articles were complemented by related literature reviews \citep[e.g.,][]{li2022towards,muurisepp2022activity} and studies \citep[e.g.,][]{netto2001socio,yabe2023behavioral}, and a few major developments in the field after the initial data collection \citep[e.g.,][]{nilforoshan2023human,xu2024experienced}. \par

\begin{table*}[!ht]
\caption{Keywords for literature search.}\label{tab:keywords}
\centering
\begin{tabularx}{0.9\textwidth}{lX}
\hline
Theme              & Keywords  \\ \hline
(1) Segregation    & segregation, spatial integration, social integration, socio-spatial integration, social cohesion    \\
(2) Activity space & spatial mobility, human mobility, daily mobility, personal mobility, individual mobility, spatio-temporal mobility, spatiotemporal mobility, socio-spatial mobility, sociospatial mobility, urban mobility, spatial movement, activity space, action space, spatial network, spatial behavior, spatio-temporal behavior, spatiotemporal behavior, use of space, lifeworld, person-based, individual-based \\
(3) Co-presence      & social mix, encountering, encounter, encounter network, social ties, third places, cross-cultural encounters, shared experiences, connectivity, co-presence, co-existence, co-presenting \\
(4) Built environment      & mobility, access inequality, accessibility, access, social and spatial inequality, transport-related social exclusion, urban sprawl, transport modes, modal split, transit deserts, transport justice, active transportation, transit-oriented development, multi-modality, travel behavior, transport affordability, traffic congestion, public transport subsidies \\
Search query & Titles, abstract, and keywords include (1 AND 2) OR (1 AND 3) OR (1 AND 4) \\\hline
\end{tabularx}
\end{table*}

Based on these studies, we first define socio-spatial segregation and propose a conceptual framework for its quantification (Section \ref{sec:measure}). 
This framework lays the foundation for reviewing empirical evidence and addressing the research questions in this study. 
We aim to offer a structured approach to socio-spatial segregation research using empirical mobility data, acknowledging the diverse and often inconsistent use of concepts in different fields. 
While we do not claim this framework as the definitive classification of methods or concepts, we hope it provides clarity and a useful basis for future studies and discussion. 
For a more thorough conceptual exploration, we refer the readers to a literature review by \cite{netto2024decoding}. \par

The covered studies are divided into two categories: a) the themes of Segregation \& Activity space, and b) the themes of Segregation \& Co-presence or Segregation \& Built environment (see Table \ref{tab:keywords}).
Category a) covers studies that employ empirical data to quantify socio-spatial segregation from an activity space viewpoint. 
They rely on traditional data sources like travel surveys, census, and register data, or those employing emerging data sources, including geolocation trackers and mobile phone data.
We reflect on the methodologies used in studies that apply emerging mobility datasets (Section \ref{sec:methods}), to answer the first research question.
Then, the studies in Category a) are synthesized to reveal the relationship between human mobility patterns and individual segregation levels, comparing residential segregation levels with the ones measured across activity space i.e, experienced segregation (Section \ref{sec:exp}). \par

The studies in Category b) align with the themes of Segregation \& Co-presence and Segregation \& Built Environment.
Their findings are synthesized to explain the relationship between mobility patterns and segregation levels and how co-presence between population groups is facilitated by the built environment (Section \ref{sec:fact_prom}). 
They contribute to proposing potential solutions to mitigate socio-spatial segregation. \par

\section{Measuring socio-spatial segregation}\label{sec:measure}
Socio-spatial segregation reflects the degree of spatial separation among different socioeconomic and demographic groups, including race/ethnicity \citep[e.g.,][]{vachuska2023racial}, birth background \citep[e.g.,][]{bertoli2021segregation}, income \citep[e.g.,][]{moro2021mobility}, education \citep[e.g.,][]{zhang2022residential}, housing \citep[e.g.,][]{zhang2019reside}, etc.
Estimating segregation requires first to identify individuals in the same location and subsequently to assess the mix of populations in and across various locations using quantitative metrics \citep{yao2019spatial, li2022towards, muurisepp2022activity}.
The metrics developed in the literature cover different aspects of the phenomenon: evenness, especially the dissimilarity index and its variants, isolation-exposure, concentration, centralization, and clustering \citep{massey1988dimensions}, which can be combined into spatial exposure/isolation and spatial evenness/clustering metrics \citep{reardon2004measures}. \par

\begin{figure*}[!htp]
\centering
\includegraphics[width=1\textwidth]{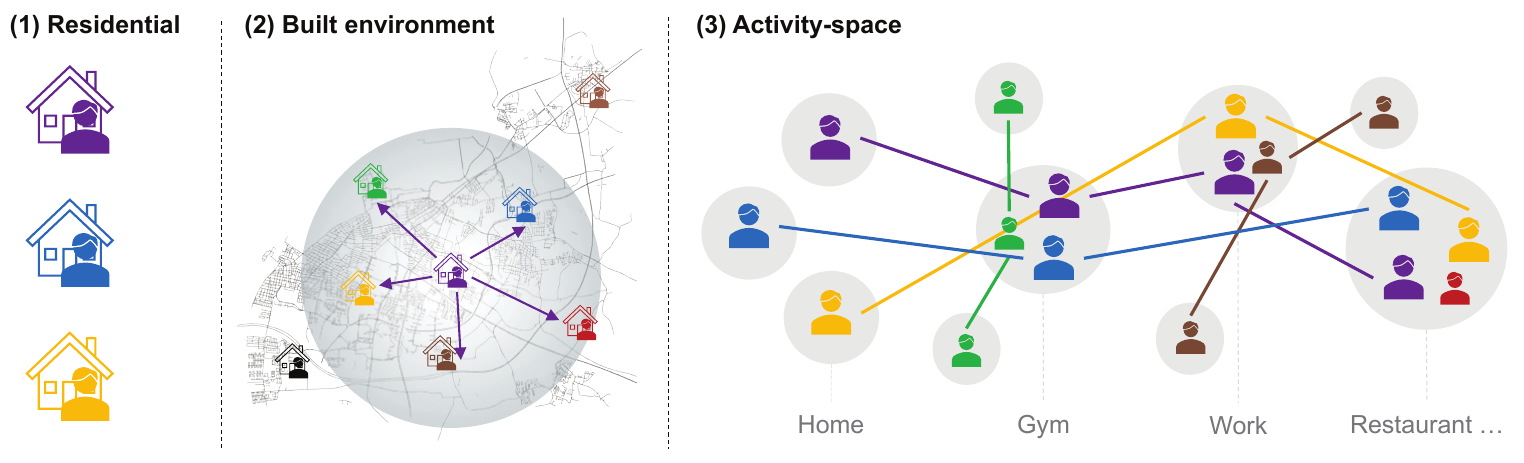}
  \caption{Conceptual framework of measuring socio-spatial segregation. (1) Residential segregation quantifies the extent to which different groups live separately from one another in different neighborhoods. (2) Built environment approach analyzes transport networks and urban spaces to evaluate the potential of reaching different groups from one's residence. (3) Activity space approaches driven by empirical mobility data assess individuals' co-presence with others across their activity space and quantify the social mixing level in these geographic areas. This study includes (2-3) for the evidence synthesis.}\label{fig:measuring_seg}
\end{figure*}

This section first presents a brief discussion around co-presence (Section \ref{sec:measure_conc}), a precursor of social interactions, and a concept used to quantify socio-spatial segregation.
Then, the section summarizes the three main approaches to measuring socio-spatial segregation (Section \ref{sec:measure_app}).
Due to limited space, we refer the reader to Section \ref{secb:measure_metr} for a more detailed introduction to the different metrics applied in each approach.
Further, we refer to the study by \cite{yao2019spatial} for a more systematic review of the metrics and models used to quantify spatial segregation.

\subsection{Co-presence: a precursor to social interaction}\label{sec:measure_conc}
A key concept for measuring socio-spatial segregation in the literature is that of \emph{co-presence}. 
In this review, we define \emph{co-presence} as the state where two or more individuals are present in the same location simultaneously. 
In other words, co-presence describes the spatial arrangement of population groups within a specific time period.
This definition originates from space syntax theory \citep{hillier2007city}, which examines the relationship between spatial configurations (e.g., road networks) and social behaviors (e.g., mixing between groups) within built environments. 
In the literature, other terms such as \emph{exposure, encounter, mixing, co-existence, and co-location} \citep{rokem2018segregation, deurloo2022co, csevik2022coexistence, juhasz2023amenity, nilforoshan2023human} are used interchangeably to indicate co-presence. 
While these terms have distinct literal meanings, they are used to describe data preparation steps for quantifying socio-spatial segregation, specifically identifying individuals who share the same space at the same time. \par

Socio-spatial segregation shows the arrangement of different groups in physical spaces, limiting interactions between these groups within those spaces.
However, co-presence itself is insufficient to quantify the social interaction level, but a necessary step for computing socio-spatial segregation.
Co-presence in urban spaces creates opportunities for different groups to meet, serving as a precursor to social interaction \citep{collins2004interaction,netto2015segregated}.
Therefore, quantifying co-presence patterns and socio-spatial segregation is meaningful in understanding and promoting social interactions.
Due to limited space, we present other key concepts within the socio-spatial segregation literature and their definitions in Table \ref{tab:concepts}. \par

A vast body of literature on socio-spatial segregation focused on co-presence within an individual's residential neighborhood.  
Recent approaches consider that individuals can also be co-present with others when outside their neighborhood. 
We see a rapidly growing number of studies leveraging observed mobility patterns of individuals to quantify co-presence empirically. 
The empirical mobility data in this literature varies, including ``small'' traditional travel surveys \citep{park2018beyond} and ``big'' mobile phone GPS records \citep{moro2021mobility}. \par

\subsection{Approaches and their spatiotemporal scale}\label{sec:measure_app}
Figure \ref{fig:measuring_seg} illustrates three primary approaches to measuring socio-spatial segregation, according to how co-presence is evaluated: the \emph{residential}, \emph{built environment}, and \emph{activity space} approaches.
Each approach provides distinct insights and operates across different spatial and temporal scales, complementing rather than forming a strict hierarchy. 
Residential areas are naturally embedded within the broader activity space, while the built environment acts as an intermediary layer that emphasizes spatial accessibility and network design, directly influencing individuals' mobility and potential co-presence. 
This framework captures both static and dynamic aspects of socio-spatial segregation, highlighting the unique contributions of each approach in advancing our understanding of socio-spatial segregation.
In this section, we discuss these three approaches. \par

Traditionally, most studies have focused on \emph{residential segregation}.
Here, segregation is considered a static area-based phenomenon \citep{duncan1955methodological}. 
The idea is to measure the co-presence of population groups within their area of residence, often an administrative or statistical subdivision \citep[e.g.,][]{andersson2010counteracting}, and then to evaluate how the population groups are spatially separated across these areal units. \par

A second approach, based on the built environment, focuses on measuring how transport networks and urban spaces can bring different populations together \citep{netto2015segregated, carpio2021multimodal,milias2024bridging}.
Most of the studies that use this approach quantify the potential opportunities for co-presence between different population groups through street network centrality measures developed within the field of space syntax analysis \citep{rokem2018segregation, carpio2021multimodal, yunitsyna2023investigating}. 
Therefore, the type of segregation measured in these studies can be defined as \textit{network segregation} for the synthesis purpose.
Similar terms are urban segregation \citep{rokem2018segregation}, mobility-aware approach \citep{carpio2021multimodal}, etc.
This type of analysis focuses on potential co-presence in locations individuals can reach from home through walk, car, or transit networks \citep[e.g.,][]{rokem2018segregation}. 
This perspective considers that individual mobility facilitates co-presence outside the residence, but the analysis methods are not based on empirical mobility data. \par

A third approach underscored in recent research conceptualizes socio-spatial segregation dynamically, by considering individuals' travel behavior and daily visited locations.
Studies based on this approach are driven by empirical mobility data from traditional sources such as travel surveys \citep{ravalet2006segregation,wang2012activity,le2017social,park2018beyond,landis2022minority,lin2023does} or emerging ones such as mobile phone GPS records and social media geolocation data \citep{osth2018spatial,wang2018urban,candipan2021residence,moro2021mobility, huang2022unfolding, wu2023revealing}.
Here, we refer to the type of segregation measured in these studies as \emph{activity space segregation}, as applied in a review by \cite{muurisepp2022activity}.
The mobility data commonly used to measure segregation for large populations include high-resolution location data from smartphone applications \citep{moro2021mobility}, telecommunications companies \citep{osth2018spatial}, and geotagged tweets \citep{netto2018temporal, wang2018urban,candipan2021residence}. 
These data can include the geolocations of millions of individuals over months and years, at the resolution of meters and seconds \citep{barbosa2018human}.
Utilizing large-scale digital data, it was shown that there is a significant correlation between co-presence and social interactions \citep{blumenstock2013social}, demonstrating that human mobility data offers a realistic approximation of co-presence between population groups \citep{nilforoshan2023human}. \par

The activity space approach describes dynamic segregation building on empirical mobility data and focusing on two aspects: urban spaces (\textit{visiting segregation}) and individuals (\textit{experienced segregation}). 
Some recent studies combine the two perspectives \citep{xu2019quantifying,moro2021mobility}. 
The term ``visiting segregation'' is from a study on income segregation using mobility data by \cite{moro2021mobility}, similar to exposure segregation defined in \cite{nilforoshan2023human}. 
The term ``experienced segregation'' was defined in \cite{moro2021mobility} and \cite{athey2021estimating}, and later widely adopted by other studies e.g., \cite{wu2023revealing} and \cite{xu2024experienced}.
The term experienced segregation is similar to multi-contextual segregation defined in \cite{park2018beyond}. \par

The stream on \textit{visiting segregation} focuses on urban spaces, considering the time-varying co-presence of different population groups \citep{nilforoshan2023human}. 
It seeks to understand how segregated a given location is, given how diverse its visitors are \citep{netto2015segregated,phillips2021social}. 
These studies examine the composition of the visitors of urban spaces based on characteristics such as income, ethnicity, birth background, etc. \citep[e.g.,][]{netto2015segregated,moro2021mobility}. 
Unlike place-based segregation \citep{kwan2009place}, e.g., workplace segregation \citep[e.g.,][]{boterman2016cocooning}, \textit{visiting segregation} considers time-varying co-presence in different urban places, rather than belonging to a workplace or a neighborhood. \par

In contrast, the stream on \textit{experienced segregation} focuses on individuals. 
It captures how much a person is co-present with diverse groups as they go about their daily lives \citep{wu2023revealing}. 
In other words, experienced segregation refers to the overall level of segregation of a person as a combined result of her residential location, travel behavior patterns, and the locations where she conducts her daily activities (activity spaces).
It considers the average segregation level they experience across these activity visits \citep{zhang2019reside,ta2021activity}. 
It is worth noting that \textit{experienced segregation}, as defined here, does not capture the actual social interactions individuals experience (see Section \ref{sec:measure_conc}). 
How co-presence with diverse individuals translates to meaningful social interactions between groups is another crucial topic in segregation research \citep{legeby2015streets}, which studies based on passively collected mobility data alone can not answer. \par

Due to data availability, experienced segregation has been commonly approximated by measuring the stays at various places over time, excluding interactions happening while individuals are moving (e.g., on public transport) \citep[e.g.,][]{moro2021mobility}. 
However, some studies particularly investigate en route segregation, highlighting the importance of equitable transport systems in reducing socio-spatial segregation during travel \citep{shen2019segregation, abbasi2021measuring, zhou2023delineating}. \par

In summary, there are different facets of measuring socio-spatial segregation (Table \ref{tab:seg_concepts}).
The literature has witnessed a clear paradigm shift from a traditional static view toward a mobility perspective that is dynamic and data-driven \citep{li2022towards}.
Big geolocation data are widely used to better understand socio-spatial segregation in the urban landscape, focusing more on individuals \citep{muurisepp2022activity}. 
Different definitions of segregation and related concepts in the literature are summarized in Table \ref{tab:concepts}. \par

\begin{table*}[!ht]
	\caption{Three approaches to socio-spatial segregation by their spatiotemporal scale of (potential) co-present individuals.}\label{tab:seg_concepts}
	\begin{center}
\begin{tabular}{lllll}
\hline
Approach & Segregation type & Perspective                  & Subject       & Time scale \\\hline
Residential & Residential  & \multirow{2}{*}{Urban space} & \multirow{2}{*}{Residents} & \multirow{2}{*}{Static (year)}          \\
Built Environment & Network  &                              &                            &                                \\\hline
\multirow{2}{*}{Activity space} & Visiting    & Urban space                  & Visitors                   & \multirow{2}{*}{Dynamic (minutes--hours)} \\
 & Experienced  & Individual                   & Travels and activities    & \\
\hline
\end{tabular}
\end{center}
\end{table*}

\section{Methodological reflections}\label{sec:methods}
Segregation studies that use extensive mobility data, e.g., collected from smartphones, can provide a rich and nuanced picture of segregation for large populations.
However, it is important to be careful when interpreting and comparing the results of these studies.
There is no consensus on the best methods for analyzing this data, and there are inherent differences in the data collection and analysis processes. 
Consequently, the findings may vary due to these methodological differences. \par

In this section, we present and reflect upon the methodologies used in the studies included in this review (45 studies and the methodological aspects of each of them are presented in more detail in the Supplementary material.)
First, we present how the term socio-spatial segregation is used differently across studies (Section \ref{sec:seg_use}), followed by how it is computed (Section \ref{sec:seg_compute}), and lastly, we discuss the limitations of existing approaches (Section \ref{sec:limits}). \par

\subsection{Use of the term segregation}\label{sec:seg_use}
The studies referenced in this review engage with the term \emph{segregation} either directly or indirectly. 
60\% of the referenced studies engage directly with the concept by trying to quantify it.
For instance, the unevenness metric has been developed to quantify visiting and experienced income segregation in the US, capturing the uneven spatial distribution of groups by income quantiles \citep{moro2021mobility}. \par

The remaining 40\% of the referenced studies only indirectly address the term segregation, i.e., without quantifying it using specific metrics. 
These studies examine disparities in the mobility patterns and activity spaces of different population groups, only hinting at the segregation of such groups across urban spaces.
For example, \cite{wu2022human} apply social media data to reveal the isolation of different racial-ethnic and economic groups in the US cities via their distinct human movement patterns.

\subsection{Discrepancy in computation}\label{sec:seg_compute}
In this section, we reflect on three key methodological aspects: the measurement of activity space (Section \ref{sec:seg_compute_act}), the choice of individuals considered co-present (Section \ref{sec:seg_compute_cop}), and the time resolution considered to measure co-presence (Section \ref{sec:seg_compute_temp}). \par

\subsubsection{Measuring activity space}\label{sec:seg_compute_act}
The term ``activity space'' refers to the set of places individuals visit as a result of their daily activities. 
It can be measured in different ways and at different spatial resolutions. 
While some studies consider the set of specific points of interest that people visit, e.g., restaurants, museums \citep{moro2021mobility}, others focus on entire administrative regions, e.g., districts \citep{Silm_Ahas_Mooses_2018}, or defined grid areas, e.g., grids of 0.6 km$^2$ \citep{zhang2022temporal}.  \par

Among the studies we reviewed, 18\% measure activity spaces with the spatial resolution of point-of-interest, while the rest use administrative units (38\%), e.g., census block groups \citep{hilman2022socioeconomic}, customized areal units (42\%), e.g., Voronoi cells \citep{moya2021exploring}, and network edges (2\%).
Notably, these different definitions of ``activity space'' can significantly influence the resulting measurements and analyses because socio-spatial segregation can occur even within relatively small areas. 
It was shown, for example, that two restaurants located next to each other can cater to distinct populations \citep{moro2021mobility}.
Hence, studies that consider co-presence over relatively large areas \citep[e.g.,][]{silm2014ethnic} due to the low spatiotemporal resolution of the data, e.g., call detail records (CDR) or social media data, may conceal detailed experienced segregation patterns in a specific location.
Geolocation records, instead, capture visits with higher granularity and thus offer a more nuanced and accurate description of spatial segregation.

\subsubsection{Measuring co-presence: Individuals included}\label{sec:seg_compute_cop}
A key methodological choice when measuring co-presence has to do with whether the visitor of a given area is regarded as co-present with the residents or with the other visitors of that area. \par

Approximately 24\% of the reviewed studies consider that any individual visiting a given area is co-present with the residents of that area. 
Conversely, 54\% of the studies consider the individual to be exposed to other visitors of that area (see more details in Supplementary material). 
The distinction is critical because the characteristics of residents \citep[e.g.,][]{wang2018urban} and visitors to an area \citep[e.g.,][]{nilforoshan2023human} can differ significantly. 
This methodological inconsistency affects how mobility patterns affect segregation (Sections \ref{sec:exp_rep} and \ref{sec:exp_red}), as these approaches capture co-presence at different levels \citep{moro2021mobility, athey2021estimating}. \par

Measuring co-presence among visitors provides a more accurate approach to quantifying activity space segregation than assessing co-presence between visitors and residents who may not stay in their residential area during the evaluation time period.
Therefore, we cover the studies measuring co-present visitors for empirical evidence synthesizing (Sections \ref{sec:exp_rep}-\ref{sec:exp_red}). \par

\subsubsection{Measuring co-presence: Temporal resolution}\label{sec:seg_compute_temp}
Another important aspect related to the measurement of segregation is the temporal resolution considered when measuring co-presence.
In principle, any two individuals should be considered co-present only if they are located in the same area at the same time (computationally, this can be approximated by considering very short time intervals). \par

However, only 31\% among the referenced studies explicitly adopt this approach and discuss how segregation varies by time of day. 
These studies measure co-presence considering time intervals that range from five minutes \citep{osth2018spatial} to day/night periods \citep{moya2021exploring}.
Nonetheless, because of the trade-off between spatial and temporal data granularity, most studies aggregate co-presence considering a period of one day to examine how co-presence differs by day of the week or weekday/weekend \citep{mooses2016ethnic}.
One recent study captures fine-scale co-presence, i.e., 50 meters of each other within 5 minutes, to identify pairs of individuals co-present with each other \citep{nilforoshan2023human}.
These efforts have advanced the field towards more accurately quantifying ``true'' co-presence in segregation research \citep{nilforoshan2023human,xu2024experienced}.

\subsection{Limits of existing approaches}\label{sec:limits}
The heterogeneity of methods in the literature on mobility and segregation can hamper the comparability of results across studies.
Moreover, this literature has a few key limitations. 
First, most of the studies analyzed in this review are predominantly correlational. 
They focus on describing the results and associating them with ambient factors.
This highlights a significant gap in the field, emphasizing the need for more causal, counterfactual-based \citep{yabe2023behavioral}, and hypothesis-based \citep{moro2021mobility} methodologies to derive robust conclusions.
This is particularly important to enhance the real-world impact of these studies, i.e., how we use the knowledge to mitigate segregation. \par

A second key limitation of the existing work is that co-presence (socio-spatial segregation) does not necessarily capture meaningful social interactions. 
This has been revealed in several studies.
\cite{schnell2014arab} show that mobility influences aspects of segregation and integration, but it cannot fully counteract societal structural stratification that leads to limited social interactions between groups. 
\cite{zhou2019between} also note that physical proximity alone does not guarantee social interactions between co-present individuals. 
\cite{dorman2020does} even find that co-presence between groups sometimes decreases the chance of having social interactions. \par

In conclusion, the results drawn from mobility-based studies on spatial segregation are influenced by methodological choices and the extent to which co-presence in these locations approximates social interactions. 
Big geolocation data of human mobility offers a nuanced understanding of socio-spatial segregation, but this type of data alone cannot fully capture the complexity of urban segregation. 
Combining quantitative and qualitative approaches will be essential for drawing robust conclusions.

\section{Individual segregation levels: residential vs. experienced}\label{sec:exp}
Each individual's socio-spatial segregation is often measured in two ways: as a resident (residential segregation) and as a visitor or traveler (experienced segregation), as defined in Section \ref{sec:measure}.
In this section, we critically assess the existing literature on whether experienced segregation across activity space is lower or higher than residential segregation and aim to clarify how individual mobility patterns influence these observed differences (Sections \ref{sec:exp_rep}, \ref{sec:exp_red}, and \ref{sec:exp_bey}), followed by a summary of the disparities between socioeconomic and ethnicity/birth background groups in their segregation levels (Section \ref{sec:exp_disp}).
We also explore how socio-spatial segregation changes considering mobility amid and after crises, e.g., hurricanes and the COVID-19 pandemic (Section \ref{sec:exp_dis}). \par

An important question is whether experienced segregation is lower or higher than residential segregation. 
To answer this question, we draw evidence from the literature by only including studies that 1) explicitly define segregation metrics, 2) simultaneously compute residential and experienced segregation levels, and 3) measure co-present visitors (Sections \ref{sec:exp_rep} and \ref{sec:exp_red}). \par

On one hand, visiting various locations beyond one's residential neighborhood could enable one to be co-present with diverse populations. 
Hence, the segregation level across one's activity space may be lower than the segregation level at one's residence \citep{alfeo2019assessing}.
On the other hand, visiting various activity locations could reflect their residential segregation levels depending on the nature of these co-present individuals, the individuals' openness to engagement, and the broader societal and systemic factors that influence social interactions and integration \citep{shdema2018social}. 
Furthermore, how people travel is often influenced by economic, demographic, and social factors, which means that not everyone has the same ability to travel outside their residential area \citep{tiznado2023unequal}. 
The literature reveals contradictory findings, indicating that moving beyond residential areas can result in measured segregation levels that are either lower or higher than residential segregation, varying widely across different individuals and groups. \par

\subsection{Studies showing similar or higher experienced segregation than residential}\label{sec:exp_rep}

Using traditional data sources such as surveys and interviews, a small body of activity space literature concludes that moving outside residential areas exhibits a similar or higher segregation level than residential segregation.
\cite{aksyonov2011social} suggest that while residential segregation in St. Petersburg, Russia, is relatively weak, experienced segregation level measured in activity space is much more prominent. 
Similarly, in Milwaukee, Wisconsin, the US, \cite{gordon2018daily} reports that racial segregation is pronounced, with distinct patterns of daily mobility among different racial groups. 
This notion of persistent segregation patterns is further supported by \cite{le2017social}, who found that the most segregated group during the night in Paris remained the most segregated during the day, indicating a strong correlation between night-time and day-time segregation. 
Consistent with these findings, studies on ethnic groups in urban China reveal that a residents' home neighborhood continues to be a strong predictor of experienced segregation in their daily activity locations \citep{tan2017examining}.
These findings are supported by a study using empirical geolocation data, specifically one-week tracking from 36 individuals combined with additional questionnaires and interviews \citep{roulston2013gps}. \par

\subsection{Studies showing lower experienced segregation than residential}\label{sec:exp_red}
In contrast, several studies have found that accounting for daily mobility outside residence shows significantly lower experienced segregation levels than approaches that measure segregation solely at residential locations.
\cite{wong2011measuring} reveal that high levels of segregation in residential spaces might be moderated by lower levels of segregation at activity locations. 
In fact, activity places are substantially more heterogeneous regarding key social characteristics than residential neighborhoods \citep{jones2014redefining,pinchak2021activity}.
A lack of co-presence with diversity (higher segregation) in the residential neighborhood may be compensated by greater workplace exposure or transport exposure \citep{boterman2016cocooning, wang2016daily,lin2023does}.
Therefore, studies have often found lower segregation levels at activity locations than at home \citep{le2017social, li2017measuring, park2018beyond, garlick2022there,fuentes2022impact}. 
Despite the value difference between the two measures, residential segregation level remains a significant and strong correlation ($r = 0.646$, $p<0.001$) with activity space-based segregation level, in a study concerning education segregation in Beijing \citep{zhang2022residential}. \par

Emerging data sources such as geotagged tweets and mobile phone data also reveal that segregation levels are generally lower when measured using activity space than using residential data \citep{grujic2019evidence,athey2021estimating,silm2021relationship,xian2022beyond,xu2022contingency}.
Research indicates that individuals from poor and black neighborhoods in the city are more mobile within the metro area than previously thought, often traveling outside their neighborhoods for work or other activities \citep{shelton2015social}. 
This trend is also seen in Sweden's metropolitan areas, where daily mobility, especially among those who frequent central places, shows lower segregation levels \citep{osth2018spatial}. 
However, this measured difference between residential and activity space is less noticeable in areas with low accessibility and mobility levels, typically located on the city outskirts \citep{osth2018spatial}. 
In Seoul, South Korea, a significantly 20\% lower segregation level measured in the daytime was observed when compared to residential segregation levels \citep{hong2020open}, a trend similarly reflected in Turkey in the context of spatial segregation between Syrian refugees and the native population \citep{bertoli2021segregation}.

\subsection{Beyond higher-lower comparisons in individual segregation levels}\label{sec:exp_bey}
Studies using both ``small'' mobility datasets, e.g., traditional travel surveys, and ``big'' emerging data, such as mobile phone geolocation records, reveal that individual segregation levels go beyond a straightforward comparison of higher or lower experienced levels than those found in residential settings.
The relationship between residential (home vicinity and sub-neighborhood) and individual segregation outside the residence is complex \citep{schnell2014arab,selim2015landscape}.
For instance, \cite{goldhaber2007model} found a weak correlation between one's residential and individual segregation levels, indicating that factors influencing each type of spatial segregation vary.
It turns out that sharing residential neighborhoods does not necessarily translate into shared routines, particularly across different socioeconomic statuses \citep{schnell2001sociospatial,browning2017socioeconomic}.
Interestingly, \cite{lin2023does} reveal that people with higher migrant exposure in their residential areas often have lower migrant exposure measured in their activity locations, and vice versa, suggesting a negative correlation between residential and experienced segregation levels. \par

Whether experienced segregation level is measured higher or lower than residential one depends largely on individuals' lifestyle, i.e., which kind of locations they visit and when during the day \citep{zhang2023human}.
Using mobile phone data, \cite{Silm_Ahas_Mooses_2018} find that places of daily activities outside home and work are less segregated, noting a surprising trend of higher workplace segregation than residential segregation, particularly in age groups 30 – 39 and above 60. 
However, \cite{silm2021relationship} observed higher segregation levels measured in residential areas than workplaces, with segregation levels during leisure activities fluctuating based on the specific activity. 
\cite{zhang2022temporal} further contribute to this understanding by using location-based service data in Beijing, revealing more pronounced segregation at workplaces than residences and a general decrease in segregation outside these typical environments. 
Additionally, the city structure and distribution of amenities emerge as a critical factor for social mixing, where diverse and unique amenities, particularly in city centers, tend to attract a mix of socioeconomic groups \citep{juhasz2023amenity}. 
A low experienced segregation level is generally associated with being in the middle of the day, away from home, such as in places like workplaces, restaurants, commercial areas, and outdoor spaces, supported by extensive geolocation data on human mobility \citep{sampson2020beyond,abbasi2021measuring,athey2021estimating,qiao2021realistic, moro2021mobility}. 
However, not everyone can equally access these places to fulfill their daily activity demand, i.e., what they plan to do in their daily lives. \par

\subsection{Differences by socioeconomic status and ethnicity/birth background}\label{sec:exp_disp}
Different population groups have different daily-life activity spaces, depending on age, race/ethnicity, and income level \citep{moro2021mobility}. 
These differences can exacerbate spatial segregation \citep{wang2012activity}.
In this section, we discuss the effect of socioeconomic status (Section \ref{sec:exp_disp_ss}) and ethnicity and birth background (Section \ref{sec:exp_disp_en}) on individual experienced segregation. 

\subsubsection{Socioeconomic status}\label{sec:exp_disp_ss}
The wealthiest and the poorest groups demonstrate contrasting mobility behaviors and activity demand \citep{aksyonov2011social, farber2012activity, osth2018spatial}, resulting in systematically different activity spaces \citep{wang2012activity}.
\cite{heringa2014individual} reveal that social status (in terms of education and income), as well as the opportunity to perform leisure activities, influences the extent of inter-ethnic contact more than neighborhood attributes. 
In turn, performing leisure activities is strongly influenced by economic factors, according to large-scale mobile phone data collected in Stockholm \citep{toger2023inequality}. \par

Table \ref{tab:disp_income_eth} summarizes the characteristics of individual mobility and activity spaces by income level observed in various studies.
Data from developed countries shows that wealthier individuals use various types of urban areas \citep{wang2022time} and travel longer distances \citep{xu2022beyond}, more likely to form connections with all classes \citep{farber2012activity}. 
In contrast, the less wealthy have limited activity spaces, leading to a more localized life. 
These may translate to the higher experienced segregation levels of less wealthy populations \citep{wu2022human}.
In Hong Kong, this mobility gap between high- and low-income is widening between 2002 and 2011 \citep{tao2020does}.  \par

\begin{table*}[!ht]
\caption{Characteristics of individual mobility and activity space by income level and ethnicity observed in studies.}\label{tab:disp_income_eth}
\centering
\begin{tabularx}{1\textwidth}{p{3cm}p{1.5cm}p{6cm}X}
\hline
Aspect & Group        & Mobility patterns     & Activity space                      \\\hline
\multirow{2}{*}{Income} & Wealthy      & Longer trip distance \citep{xu2022beyond,farber2012activity} & Diverse and widely spread \citep{wang2022time} \\
& & & Sports, leisure \citep{zambon2017beyond}, business \citep{wang2012activity, heringa2014individual}, shopping \citep{aksyonov2011social} \\
& Less wealthy & Shorter trip distance \citep{wu2022human} & Constrained and localized \citep{netto2015segregated, zhou2015social}           \\
&             & Less frequently travel outside city \citep{aksyonov2011social} & Convenience stores \citep{aksyonov2011social} \\\hline
\multirow{2}{*}{\makecell{Ethnicity/\\birth background}} & Majority & Longer commuting \citep{garlick2022there}        & Social, recreational, errand \citep{shirgaokar2021differences} \\
& Minority & Shorter travel distance \citep{silm2014ethnic} & Exercise, education \citep{shirgaokar2021differences} \\\hline
\end{tabularx}
\end{table*}

In some contexts, however, a different effect was observed, with the wealthy living relatively more segregated lives \citep{shelton2015social, xu2019quantifying} than the less wealthy.
\cite{atkinson2016limited} illustrate that the super-rich in London create a ``cloud space'' or ``flowing enclave,'' engaging with the city's diversity in a limited way. 
This phenomenon of spatial segregation among the upper classes is also observed in the Paris region, where \cite{le2017social} note that they remain the most segregated group both in residence and during daytime activities.
In developing countries, the less wealthy population does not always travel less or have more limited activity spaces than the rest \citep{wissink2016bangkok,moya2021exploring}. \par

Socio-economic status often interacts with housing type, which affects individual mobility and experienced segregation level \citep{demoraes2021live}.
This is evidenced by the poor families in public housing facing increased isolation \citep{li2017measuring} and varied activity spaces among social groups \citep{zhang2019reside}, resulting in limited interaction opportunities between groups.

\subsubsection{Ethnicity and birth background}\label{sec:exp_disp_en}
Ethnic groups exhibit distinct experienced segregation levels \citep{raanan2014mental}.
These disparities of measured segregation levels by ethnicity or birth background can be ascribed to their distinct mobility and activity space patterns (Table \ref{tab:disp_income_eth}).
For example, the study by \cite{jarv2015ethnic} suggests that ethnic differences in spatial behavior become more pronounced in leisure-related activities and other non-routine behaviors \citep{silm2014ethnic}.
Ethnicity also affects transport choices. 
Living in co-ethnic neighborhoods increases the likelihood of inter-household carpooling for Asian and Hispanic groups, while this is not the case for African Americans \citep{shin2017ethnic}. \par

Intersectionality has received attention in a few studies, as ethnicity or birth background often intertwines with factors like income and education. 
High-income natives \citep{boterman2016cocooning} and low-income minorities \citep{tan2019social} can show higher experienced segregation levels than the other populations.
Most racial-ethnic groups' ethnicity/birth background segregation increases along with higher economic status \citep{wu2023revealing}, except for Asian groups with diverse interactions regardless of economic status \citep{salgado2021street}.

\subsection{Amid and post-crisis segregation levels}\label{sec:exp_dis}
Socio-spatial segregation can change dramatically during and after crises.
This section reviews the findings in the field, focusing on the impact of COVID-19 and natural disasters, where we observe a surge in the use of extensive mobile phone geolocation data from large populations in understanding socio-spatial segregation.
These studies provide valuable insights into segregation and related policy interventions.
Instead of focusing solely on general findings, this section complements the review with crisis-specific insights related to segregation. \par

The COVID-19 pandemic tends to amplify segregation.
In the US, racial segregation significantly influenced urban spatial patterns and behaviors, such as public transportation usage \citep{hu2022examining}, further intensifying segregation \citep{marlow2021neighborhood, li2022aggravated, lu2023people}. 
Similarly, in Sweden, the pandemic exacerbated socio-economic and ethnic segregations, resulting in high mortality rates in low-income, multi-ethnic neighborhoods \citep{joelsson2022cracks}. \par

By comparing pre- and amid-pandemic data, the literature highlights that changes in mobility patterns, such as reduced public transport usage and increased reliance on cars, have deepened existing inequalities \citep{shin2021spatial,bonaccorsi2021socioeconomic}. 
According to amid-pandemic observations and evidence synthesis, this impact is particularly pronounced among young and vulnerable groups in socially disadvantaged neighborhoods, who often depend on public transport for their jobs with irregular hours \citep{joelsson2022cracks}. 
Similarly, in cities like New York, Los Angeles, and Sao Paulo, poor peripheries with high population density and limited access to individual transportation options faced heightened virus transmission risks \citep{sathler2022city}. \par

Besides pandemics like COVID-19, segregation is a salient issue in evacuations during and after disasters.
Large populations are displaced in this context, underscoring social inequalities and intensifying segregation.
Analyzing large-scale mobility data, \cite{yabe2020effects} found that higher-income individuals were more likely to evacuate disaster-affected areas and relocate to less damaged areas. 
This disparity in mobility resulted in prolonged spatial income segregation post-disaster, with higher-income individuals avoiding severely damaged areas while lower-income individuals remained, exacerbating segregation. 
Similar patterns were observed during Hurricane Harvey's evacuation \citep{deng2021high}. \par

In summary, crises tend to magnify existing segregation, mainly through individual differences in mobility patterns. 
Given the critical role of mobility in disaster evacuation and disease transmission, it is imperative to closely examine the mobility disparities across socioeconomic groups and formulate targeted policies to reduce segregation and mitigate the adverse effects of disasters.

\section{Explaining experienced segregation}\label{sec:fact_prom}
Using activity space approaches on empirical mobility data, studies have revealed persistent segregation in individuals' daily mobility and shed light on how different populations exhibit various levels of segregation in their daily lives (Section \ref{sec:exp}). \par

\begin{figure*}[!htp]
\centering
\includegraphics[width=0.6\textwidth]{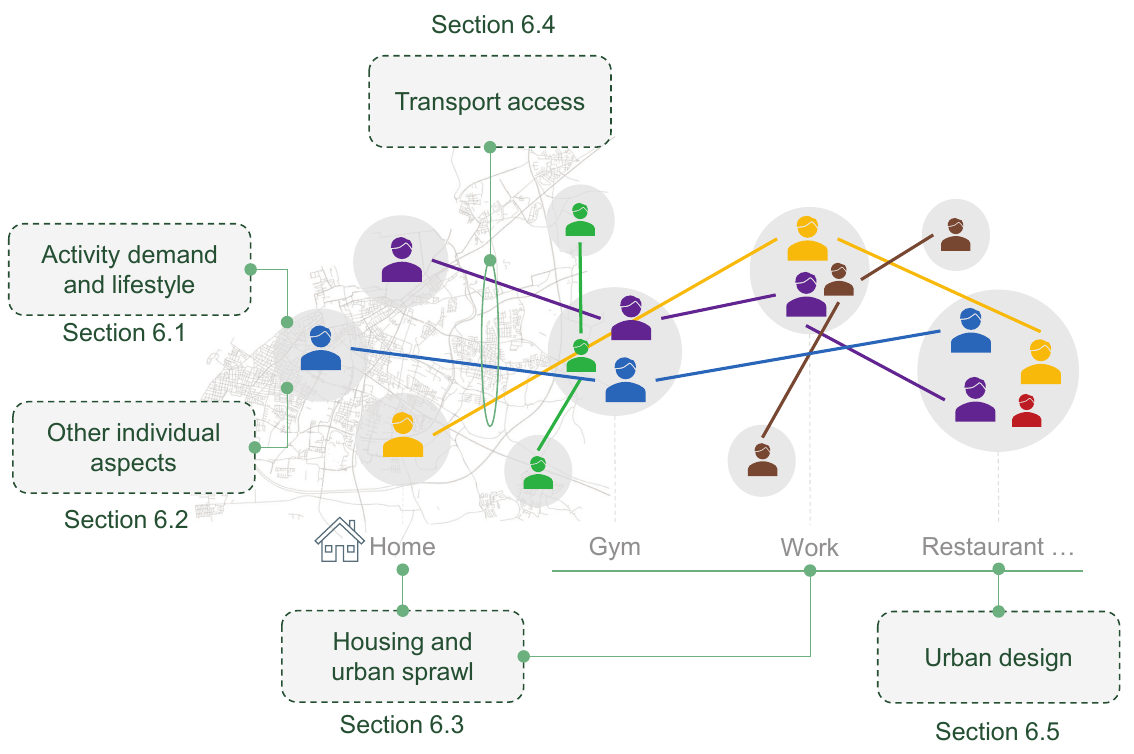}
  \caption{Conceptual framework explaining socio-spatial segregation. Activity demand and lifestyle (Section \ref{sec:exp_disp_homo}), and Other individual aspects (Section \ref{sec:exp_disp_o}) deal with individual factors. Housing and urban sprawl (Section \ref{sec:exp_seg_h}) examines housing and its relationship with workplace locations. Transport access (Section \ref{sec:exp_seg_t}) refers to access from one's residence, linking the home with the rest of the activity space. Urban design (Section \ref{sec:exp_seg_ud}) covers the spatial aspect of activity space.}\label{fig:explaining_seg}
\end{figure*}

In this section, we further integrate the insights from the studies covering the themes of segregation, co-presence, and built environment to systematize key factors that help explain observed disparities in individual segregation levels (Figure \ref{fig:explaining_seg}).
Individuals' experienced segregation is related to both subjective factors, e.g., preferences for certain activities \citep{moro2021mobility}, and objective factors, e.g., unequal access to diverse social settings \citep{netto2015segregated} such as housing, transport access, etc.
These factors contribute to different levels of mobility and shape individuals' activity spaces. 
Ultimately, they translate into distinct movement networks, leading to different co-presence opportunities.
These daily experiences eventually create homophilic personal networks, therefore perpetuating segregated class networks.
Taken together, these elements shape how individuals are co-present with different population groups. \par

Covering various components of how socio-spatial segregation is produced across activity space, we investigate five aspects of explaining experienced segregation in this section: activity demand and lifestyle (Section \ref{sec:exp_disp_homo}), other individual aspects such as security, neighborhood trust, etc. (Section \ref{sec:exp_disp_o}), housing and urban sprawl characterizing the relative spatial relationship between residence and other activity locations (Section \ref{sec:exp_seg_h}), transport access describing how easy one can reach various resources and opportunities (Section \ref{sec:exp_seg_t}), and urban design characterizing the built environment of activity spaces (Section \ref{sec:exp_seg_ud}). \par

\subsection{Activity demand and lifestyle}\label{sec:exp_disp_homo}
Individuals are more likely to interact with others from similar backgrounds, a phenomenon called homophily \citep{amini2014impact,xu2022beyond}.
This can happen along many socioeconomic dimensions, such as income \citep{yip2016exploring,wu2022human}, race/ethnicity \citep{gordon2018daily,jones2014redefining,bora2014mobility,hedman2021daily}, birth background \citep{hedman2021daily}, and religion \citep{davies2019networks}, even when other factors are controlled such as traveling distance \citep{wang2018urban}, travel time, and opportunities for interaction \citep{heine2021analysis}.
In short, people are less likely to travel between areas considered to be distant not only in geographical terms but also in socioeconomic terms \citep{park2021we,chen2024behavioral}. \par

Individuals' activity demands and lifestyles partly drive this homophily, leading to their different experienced segregation levels.
Studies based on GPS data have revealed that experienced segregation is affected by individuals' lifestyles, as captured by the kind of places that they visit in their daily life \citep{moro2021mobility}.
Lifestyles with more socializing, shopping weekends, and coffee shop visits are associated with higher social integration \citep{yang2023identifying, nilforoshan2023human}. 
Instead, individuals visiting entertainment venues and restaurants may show higher levels of experienced segregation because those locations cater to specific income groups, and their cost or cultural context may exclude lower-income individuals \citep{moro2021mobility}. 
\cite{di2018sequences} identify that depending on spatial spending records, i.e., the records of where money is spent on what, a strong homophily effect can be observed in a few particular groups, such as commuter, household, young, hi-tech, dinner-out, and average lifestyle groups.
\cite{moro2021mobility} suggest that people who actively explore different places tend to be more economically integrated, showing a lower level of experienced income segregation. \par

The evidence reveals that individuals residing in the same area with similar housing and transportation accessibility may have different experienced segregation levels due to their diverse activity demands and lifestyles \citep{yang2023identifying}.
Lifestyle changes such as working from home, e-commerce, food delivery, etc., can affect individuals' experienced segregation levels.
For example, working from home increases isolation \citep{xiao2021impacts} and limits the interactions in the residential neighborhood, contributing to a high level of experienced segregation.
Counterfactual analysis by \cite{yabe2023behavioral} revealed that, two years after the first COVID-19 wave, changes in experienced segregation persist mainly due to lifestyle changes, such as reduced willingness to explore new places.

\subsection{Other individual aspects}\label{sec:exp_disp_o}
Besides activity demand and lifestyle, individual values, fears, trust, and social networks affect individuals' experienced segregation level.
\cite{jarv2021link} identify the subjective self-estimated social status as a critical factor affecting the extent of activity spaces and experiences of segregation.
People's inclination towards ascriptive (traditional roles) or achievable (personal accomplishments) status affects their spatial segregation \citep{goldhaber2007model}.
Those focusing on achievable status are more likely to move beyond their ethnic enclaves, while ascriptive-oriented individuals stay within them. \par

Security barriers and fear also affect the ability to move, interact, and use specific spaces, perpetuating the cycle of segregation \citep{roulston2013gps, selim2015landscape}.
\cite{dixon2020parallel} suggest that religious segregation results not only from socioeconomic and institutional forces but also from individual mobility choices influenced by perceived intergroup threats and contact experiences. 
This perceived fear often restricts residents' willingness to travel, leading to geographical isolation and limited access to work, education, and other activities \citep{hernandez2016mobilities}. 
Furthermore, trust plays a crucial role in individuals' experienced segregation, with \cite{browning2017socioeconomic} finding that high levels of neighborhood trust can mitigate the effects of socioeconomic inequality on spatial segregation in daily routines. \par

Besides individual values, fears, and trust, social networks within ethnic groups strongly affect segregation patterns \citep{silm2014ethnic}.
Segregation experiences can vary significantly between individuals; they show a strong correlation at a broader social level between social network segregation and spatial segregation \citep{xu2019quantifying}.

\subsection{Housing and urban sprawl}\label{sec:exp_seg_h}
Housing locations and types primarily affect individual mobility and experienced segregation (Table \ref{tab:housing}).
Minorities often live in disadvantaged neighborhoods, which makes it hard to be co-present with the other groups \citep{tao2020influence}, with the built environment's slow evolution locking urban communities into persistent settlement patterns and inequalities to resources \citep{patias2023local}. \par

\begin{table}[!ht]
\caption{Housing effects on experienced segregation.}\label{tab:housing}
\centering
\begin{tabularx}{1\textwidth}{p{4.7cm}X}
\hline
Aspect                            & Impact                                                                                                                                        \\\hline
Rural vs. urban                   & Rural migrants have higher birth background segregation levels in both residential and activity spaces than urban migrants \citep{lin2023does,shen2023linking}.                            \\
Land-use diversity & Areas with high land-use diversity shows less daytime segregation, despite high levels of nighttime segregation  \citep{fuentes2022impact}.                              \\
High land values                  & Lead to segregation both day and night, attracting high-class residents and contributing to segregation  \citep{fuentes2022impact}.                                      \\
Disadvantaged neighborhoods       & Face difficulties mixing with other groups due to persistent settlement patterns,  with urban evolution fostering segregation into "ghettos" \citep{power2012social}. \\
Peripheral and disconnected areas & Poorer populations residing in these areas face socio-economic disadvantages due to limited access to services and job opportunities \citep{kronenberger2017configurational, atuesta2018access}.\\\hline
\end{tabularx}
\end{table}

Urban sprawl is the uncontrolled expansion of low-density urban areas into the surrounding rural land.
Escalating segregation levels usually accompany such increasing size of cities \citep{monkkonen2018urban,nilforoshan2023human}, observed in several countries, such as Brazil \citep{bittencourt2021cumulative}, China \citep{zhao2013impact}, Iran \citep{azhdari2018exploring}, and Chile \citep{Figueroa_Margariota_Mora_2019}.
Increasing urban compactness counteracts urban sprawl, significantly enhancing upward mobility through better job accessibility and indirectly mitigating poverty segregation \citep{ewing2016does}. 
Conversely, urban sprawl contributes to disparities in public transportation and job opportunities \citep{bittencourt2021cumulative, zhao2013impact}, a decline in street network accessibility \citep{Figueroa_Margariota_Mora_2019}, and the isolation of peripheral areas \citep{azhdari2018exploring, Figueroa_Margariota_Mora_2019}.
These aspects intensify segregation through housing aspects. \par

As a consequence of residential segregation, ``spatial mismatch'' refers to the geographical separation between low-income communities, often inhabited by racial or ethnic minorities, and employment opportunities, typically located in areas located far from these communities \citep{kain1968housing}.
Spatial mismatch creates a complex interplay between where people live and where they can work or access services.
Minority groups tend to have higher spatial mismatch levels than their white counterparts living in the same Metropolitan Statistical Areas in the US \citep{easley2018spatial}.
Spatial mismatch leads to longer commuting times for low-wage workers \citep{blazquez2010commuting, wang2022time}. 
In China, the low-income population faces a trade-off between service accessibility and the floor area of their residences \citep{chen2019accessibility}.
However, the availability of low-rent housing within urban villages, coupled with short commuting times, alleviates the spatial mismatch for disadvantaged groups in specific regions \citep{chen2021spatial}. \par

Housing and urban sprawl largely affect individuals' home and work locations, determining access to key urban resources, e.g., transport.
They are vital in shaping individuals' mobility patterns and spatial segregation levels across their activity spaces.
Housing segregation has been extensively studied within urbanization history \citep[e.g.,][]{park2019city} and understood with a variety of models \citep[e.g.,][]{schelling1971dynamic} that relate residential segregation patterns to the progressive growth of cities.
Many theories have been developed to explain observed residential segregation patterns, covering human-ecology-inspired segregation theory, behavioral theory, structural theory, political theory, and local urban and housing policy \citep{Musterd_2020}.
Due to limited space, we refer interested readers to the book by \cite{Musterd_2020}.

\subsection{Transport accessibility}\label{sec:exp_seg_t}
This section first reviews evidence on how transport accessibility differs by income and ethnicity (Section \ref{sec:exp_seg_tf}) and brings attention to transport equity and how it affects socio-spatial segregation (Section \ref{sec:exp_seg_te}). \par

Transport accessibility plays a crucial role in shaping experienced segregation patterns via mobility. 
Because spatial accessibility reflects how easy it is to reach locations and activities, it ultimately determines the range of places and social environments people can visit. 
Limited transportation opportunities are closely linked to an increased risk of social exclusion for various individuals or groups \citep{lucas2016transport, luz2022understanding}, which is accentuated in a pre- vs. post-pandemic analysis by \cite{gallego2023social}. 
Given these challenges, to foster better social integration, \cite{rokem2019geographies} argues for substantially reevaluating transport infrastructure accessibility, which is central to measuring the equity impacts of transport investments \citep{pereira2017distributive}. \par

\subsubsection{Factors of income and ethnicity/birth background}\label{sec:exp_seg_tf}
Transport access disparity between low and high-income groups may explain their distinct mobility patterns, as reviewed in Section \ref{sec:exp_disp_ss}.
High-income groups generally have better accessibility than the other groups \citep{jang2021imbalance, arellana2021urban}.
Lower-income households limit where they go due to low affordability and poor accessibility \citep{logiodice2015spatial, hernandez2016mobilities, martinez2018creating,cortes2021spatial}, leading to socioeconomic segregation in their activity spaces \citep{hu2017commuting, pena2022dots, cromley2023examining}. \par

How transport access affects mobility is also associated with ethnicity/birth background.
Racial residential segregation is associated with lower equitable travel across neighborhoods \citep{haque2016discriminated} and fewer visits to common hubs \cite{sampson2020beyond}. 
In the US, living in racially segregated metropolitan areas leads to longer travel times for Black individuals when compared to White individuals, particularly when driving \citep{landis2022minority}. 
Although homophily in activity spaces holds across races, for White residents, it is more a preference rather than accessibility \citep{vachuska2023racial}. 
While for Black and Hispanic residents, this is mainly due to lower accessibility of various transport modes. 
Ethnicity/birth background often interacts with socioeconomic status, resulting in different accessibility patterns \citep{xiao2021spatial}.
\cite{rokem2019geographies} reveal two distinct migrant groups - economic migrants and refugees, where the economic migrants have activity spaces closer to the city center, indicating better mobility opportunities than refugees. \par

The poor transport access where low-income and minority groups live tends to create a vicious circle of segregation and inequalities \citep{bittencourt2021cumulative}. 
Extended travel times and restricted access to resources expose the low-income to a greater risk of long-term unemployment \citep{korsu2010job}, diminishing their opportunities for upward mobility. 
In Turkey, Syrian refugees have limited accessibility to various transportation methods and activities due to a lack of language skills, further compounding their poverty and social exclusion \citep{ozkazanc2021transportation}. 
In contrast, efficient transportation networks, e.g., better job accessibility, can offset the adverse effects of residential segregation and narrow the income disparity between socioeconomic groups \citep{galaskiewicz2021minority,eom2022does}. \par

\subsubsection{Transport equity}\label{sec:exp_seg_te}
Transport access discrimination against certain social groups partly stems from planning stages being intentionally or unintentionally biased towards privileged groups and might perpetuate existing social segregation \citep{govender2012segregation,golub2013race,naranjo2016impacts}.
For example, road infrastructure enhancements in peri-urban communities often improve connectivity and attract middle-to-high-income individuals seeking enhanced services \citep{adugbila2023road}.
However, this influx tends to displace low-income residents, pushing them into hinterlands and leading to fragmentation within these peri-urban areas.
Elevated highways and urban freeways deepen socio-spatial divisions by favoring affluent commuters and exposing marginalized groups to environmental risks and poor services \citep{graham2018elite, mahajan2023highways}.
While public transport infrastructure often supports various goals, including equitable mobility, accessibility, and affordability, its design can also inadvertently influence segregation patterns.
High-income neighborhoods benefit more from public transport investments than low-income ones, reinforcing residential segregation \citep{heilmann2018transit}. 
Compared with lower-income homeowners, higher-income homeowners can better take advantage of transit-induced price capitalization effects on their property values and upgrade to more affluent neighborhoods \citep{nilsson2020impact}.

\subsection{Urban design}\label{sec:exp_seg_ud}
Segregation is influenced not only by current socioeconomic variables but also by historical patterns of urbanization and transformation \citep{zhou2015social}.
Specific forms of urban areas are better than others in facilitating individual mobility and fostering social inclusion \citep{goldblatt2015relationship}, as illustrated in Table \ref{tab:urban_design}. \par

\begin{table*}[!ht]
\caption{Urban design effects on experienced segregation.}\label{tab:urban_design}
\centering
\begin{tabularx}{1\textwidth}{lX}
\hline
Aspect                            & Impact                                                                                                                                        \\\hline
Hierarchical urban structures & Limit public space use and social inclusion, particularly affecting immigrants \citep{legeby2011does}. \\
Distribution of consumption spaces & Drives segregation in public space usage, indicating urban exclusion \citep{bolzoni2016spaces}. \\
Streets with slower traffic and good environment & Enhance social integration and contribute to vibrant public life \citep{sauter2008liveable}. \\
High spatial integration & Leads to enhanced access to services, reducing segregation \citep{van2015ethnic,garnica2019spatial,legeby2022towards}. \\
High pedestrian density, walkability, and bikability  & Aligns with lower segregation through synergistic urban patterns with the surrounding cities \citep{alghatam2019generic,van2017subjective,mouratidis2020built, wang2022neighborhood}. \\
Multi-scalar spatial configurations & Facilitates urban encounters and improves social integration \citep{csevik2022coexistence}. \\
Mixed-use design and linear parks & Enhances activity interactions and socioeconomic integration \citep{gao2023socio}. \\\hline
\end{tabularx}
\end{table*}

Architecture, urban design, and planning are vital in mitigating inequalities, through the distribution and accessibility of resources \citep{legeby2022towards}. 
Accessibility disparities contribute to urban segregation, and targeted urban design interventions could address these inequalities, ultimately supporting more equitable cities.
Consequently, the built environment's impact on segregation variation demands careful consideration. 
To effectively address socio-spatial segregation dynamics, it is essential to examine the co-presence of social groups in public spaces, utilizing urban form to foster positive change \citep{miranda2020shape}.
As a crucial public space, urban parks enhance social integration \citep{Samson_2017}, demanding inclusive design to encourage shared activities to foster interaction among diverse community members \citep{abdelmonem2015search,xiao2019exploring}.
Providing equitable park access is crucial for mitigating experienced segregation \citep{van2015ethnic,miller2019park}.

\subsection{Facilitating mobility to promote integration}\label{sec:exp_seg_f}
Segregation measured across broader activity spaces in people's daily mobility is generally lower than residential segregation (Section \ref{sec:exp}).
Evidence in built environment research also highlights the essential role of mobility in promoting integration \citep{camarero2019thinking,mooses2020ethno,huang2022unfolding}. 
Therefore, we have this section to briefly summarize practices and policies from various regions focusing on housing, public transport, and urban design that impact segregation (detailed in Table \ref{tab:actions}).\par

Housing policies could facilitate interactions between socioeconomic groups and promote social integration when they account for affordability and lack of transport access faced by minorities \citep{Utzig_2017}.
In transport and urban development, the car-centric urban development is often associated with increased segregation \citep{sanchez2004inequitable}, restricting transport access to essential opportunities like jobs, education, and healthcare \citep{sanchez2004inequitable,mcdonagh2006transport}. 
Better public transportation may enhance co-presence levels between population groups and reduce overall levels of experienced segregation \citep{wong2011measuring,kryvobokov2014willingness, landis2022minority, power2012social, Utzig_2017,carpio2021multimodal,athey2021estimating,huang2022unfolding}.
Transit-oriented development planning is a way to break the vicious cycle of car-centric mode, where we need strategic methods to counter spatial inequality to fully unleash its potential in promoting integration in modern cities. 
In this process, \cite{mueller2018methods} emphasize the integration of affordable housing preservation into city planning, particularly near transit corridors. \par

In urban design, targeted initiatives in walkability, public space, and housing design can enhance urban vitality and social cohesion.
In implementing these initiatives, \cite{unceta2020socio} highlight the importance of considering local context and potential in space and society rather than solely relying on land regularization and imported solutions.
One example is a measuring tool by \cite{alipour2021assessing} for assessing social sustainability indicators, covering factors like density, land use, mobility options, street layouts, etc., potentially revealing segregation in public urban spaces. \par

\section{Discussion}\label{sec:discussion}
This review article makes three contributions: a conceptual framework, critical methodological reflections, and cross-disciplinary insights. 
Firstly, it defines core concepts and clarifies their connections, addressing the diverse and inconsistent terms used in the field. 
Secondly, it reviews studies that adopt activity space approaches and critically examines the methodologies employed with emerging mobility data sources. 
Lastly, by incorporating insights from broad disciplines, this review enhances the understanding of socio-spatial segregation at the individual level and offers actionable insights for reducing segregation in an interdisciplinary context. \par

Our analysis of over 170 research works reveals that increased mobility enables individuals to be co-present with more diverse populations, particularly outside their residential neighborhoods, offering the potential to reduce the experienced segregation level. 
However, the extent to which experienced segregation is lower than residential segregation for a given individual depends on various factors.
These include the nature of the encounters, the individuals' willingness to engage with others, and the broader societal and systemic influences at play. 
This interplay between residential segregation and individually experienced segregation patterns is intricate, as economic, cultural, physical, and spatial factors not only shape individual mobility patterns but also, in turn, influence individual segregation levels, both at residence and across activity space. 
Such complexity underscores the multifaceted nature of spatial segregation and its challenges. \par

This section synthesizes the literature review to summarize our answers to the research questions proposed in this study regarding people's movement in space and time and their segregation levels.
Initially, we evaluate the burgeoning literature utilizing emerging data sources, assessing its potential and challenges in understanding socio-spatial segregation (Section \ref{sec:dis_q3}).
Subsequently, we compare residential with experienced segregation levels and how human mobility relates to such disparity (Section \ref{sec:dis_q1}).
Finally, we pinpoint existing research gaps and propose corresponding future directions.

\subsection{Emerging data sources: nuanced understanding at scale}\label{sec:dis_q3}
We have found over 70 reference studies using geolocation information from emerging data sources such as GPS tracking devices, mobile phone GPS, Call Detail Records, and geotagged tweets to study socio-spatial segregation.
Studies using emerging data sources, compared to those relying on traditional data, provide evidence based on actual behavioral data on mobility rather than stated preferences and offer broader spatial and population coverage with high spatiotemporal granularity \citep{nilforoshan2023human}. \par

\subsubsection{Novel insights}
These emerging-data studies typically quantify multi-dimensional aspects of socio-spatial segregation, simultaneously examining race/ethnicity, birth background, income, and other factors \citep[e.g.,][]{heine2021analysis}. 
In contrast, traditional data-driven studies often focus on a single aspect of segregation, e.g., housing \citep{wang2012activity}, linking it to other relevant and readily available explanatory variables in the applied data. 
With data from a wide range of users, these sources offer diverse demographic insights, capturing the experiences and behaviors of different population groups and revealing interacting effects between various dimensions, e.g., income and birth background \citep{gao2021segregation}. \par

Emerging data sources often provide large-scale insights into segregation levels while preserving nuanced understandings to zoom in at the block level \citep{moro2021mobility, nilforoshan2023human}.
Most studies examining the entire country \citep{vachuska2023racial} or major metropolitan areas \citep{moro2021mobility,wu2022human,huang2022unfolding} use data collected from the US.
Leveraging these widely available data sources allows for a detailed examination of various points of interest \citep{moro2021mobility} and regions, uncovering segregation patterns at the block level \citep{wu2022human}. 
They also enable near real-time analysis with high temporal granularity, providing insights into dynamic segregation patterns and trends as they change over different times of the day \citep{shen2023linking} and seasons of the year \citep{mooses2016ethnic}. \par

The abundance and intricacy of big data have spurred the development of innovative analytical methods and tools in socio-spatial segregation research. 
For example, the theory of mobility homophily, confirmed across multiple regions \citep{xu2019quantifying,heine2021analysis,huang2022unfolding}, extends its relevance beyond residential segregation to include activity space segregation patterns. 
Leveraging emerging data sources allows for integrating extensive mobility geolocation data with social network data, unveiling patterns previously unobservable at such refined scales \citep{xu2019quantifying,silm2021relationship}. 
Building on this, \cite{moro2021mobility} have developed a social exploration and preferential return model using vast human mobility geolocation data, effectively capturing the dual aspects of economic integration: social and place explorations.
By harnessing extensive mobility data from large populations, it becomes feasible to construct a large-scale mobility network \citep{nilforoshan2023human} and employ network analysis tools, like community detection, to uncover insights into segregation patterns at a national level \citep{huang2022unfolding}. \par

This review considers segregation as a physical and spatial in-person process. 
However, emerging data sources allow us to go beyond this boundary \citep{ye2021spatial}.
In developed countries, online social media is becoming more prevalent, sometimes substituting in-person interactions. 
Thus, we only observe part of the experienced segregation from mobility data while missing the other part in the digital world. 
For example, using credit card transactions and Twitter mentions, a study found that offline and online segregation experiences are very similar in Turkey \citep{dong2020segregated}.

\subsubsection{Challenges}
Studies using emerging data sources, such as mobile phone applications and GPS-enabled services, face four significant challenges: population bias, uneven sampling of locations, association with census data, and methodology for quantifying socio-spatial segregation. \par

Population bias arises because the user demographics behind this big geodata, including age, gender, and ethnicity, often do not accurately represent the broader population. 
Therefore, appropriately weighting individual devices in spatial segregation analysis is crucial to prevent inaccurate results \citep{wang2018urban,liao2024uneven}.
\cite{saxon2021empirical} show that spatial models using mobile phone GPS data have systematic bias, notably towards overestimating the park access of minority populations, which results in understating inequity. \par

An uneven sampling of locations brings two types of biases in the collected geolocation data on human mobility. 
Firstly, self-reported geolocations such as geotagged tweets have selective biases, overly representing leisure and non-routine activities \citep{liao2019individual}.
Secondly, passively collected geodata from call detail records or GPS-enabled phone applications are contingent on user interactions with mobile phones, resulting in data sparsity and biases, particularly towards activities in the afternoon and nighttime \citep{liao2021understanding}.
These factors significantly influence the analysis of segregation experiences, considering the recorded activity spaces visited instead of the full range. \par

Analyzing socio-spatial segregation with anonymized mobile phone data necessitates implementing home detection methods to infer the sociodemographic attributes of the device users.
Traced by human circadian rhythms, temporal rules are commonly used to infer individuals' home and work locations from mobile phone data \citep[e.g.,][]{gao2021segregation}. 
There is a notable lack of validation against ground truth data, mainly due to privacy concerns and the anonymization of big geolocation data \citep{verma2024comparison}.
However, accurately identifying home and work locations remains crucial for understanding daily mobility patterns and capturing individual segregation experiences.
More work is needed to examine and enhance the validity of these methods for practical applications in socio-spatial segregation research \citep{pappalardo2021evaluation}. \par

Another limitation is regarding the collection and analysis of demographic and household data. 
The inability to directly ascertain gender, age, or family composition, alongside reliance on residential location to infer income, restricts our understanding of demographic, social, and economic diversity. 
Additionally, the focus on individual mobile devices, without access to household-level information, limits insights into how household structure affects mobility and socio-spatial segregation patterns, highlighting a significant gap in analyzing collective household dynamics. \par

The fourth challenge is the methodology for analyzing socio-spatial segregation, described in Section \ref{sec:methods}.
Studies that try to measure segregation levels using an activity space approach present significant differences in how they define activity spaces, co-presence, and the temporal resolution deployed in the analyses. 
For instance, considering co-present visitors rather than exposed residents results in more heterogeneous experienced segregation levels \citep{xu2022contingency}. 
A critical limitation is the absence of causal analysis, which confines the studies to descriptive and correlational assessments, thereby reducing their potential real-world impact.
Untangling to what extent experienced segregation results from different factors (urban design, housing, transportation, and individual aspects) remains a challenge. \par

\subsection{Residential vs. experienced segregation levels}\label{sec:dis_q1}
This study investigated the literature using the activity space approach and empirical mobility data in quantifying socio-spatial segregation to examine whether moving outside one's residence contributes to lower segregation levels than traditional residential segregation.
The literature using either traditional or emerging data sources reveals mixed results. \par

Several studies underscore the correlation between residential and experienced segregation levels, drawing attention to the phenomenon of mobility homophily.
Generally, an individual's segregation level across their activity space is lower than that measured in their residential area. 
However, these two aspects are not necessarily contradictory to each other.
Some studies highlight both sides \citep[e.g.,][]{Silm_Ahas_Mooses_2018,zhang2022temporal,lin2023does}, drawing attention to the complexity of the relationship between residential and experienced segregation levels.
Such complexity comes from different mechanisms of spatial segregation from an activity space perspective. \par

\textbf{Economic and social influences}. 
Job opportunities, leisure options, and cultural preferences based on ethnicity, birth background, and income level significantly shape individual mobility and, consequently, experienced segregation. 
Differences in wealth lead to distinct mobility patterns. 
In developed countries, those with higher socioeconomic status often engage in more diverse activities, accessing various locations, which dilutes their experienced segregation levels. 
Conversely, lower-income individuals typically have more localized mobility, intensifying their experienced segregation levels. 
In developing countries, the opposite effect has been observed. 
Different ethnic groups display unique mobility patterns, often gravitating towards or remaining within areas predominantly occupied by their communities. 
This tendency is influenced by various factors such as economic constraints, security perceptions, and targeted policing practices. \par

\textbf{Social networks and lifestyles}.
The nature of social networks within ethnic communities, neighborhood trust levels, and preferences for social interaction contribute to forming socio-spatial segregation patterns. 
Mobility homophily, or the tendency to interact with similar population groups, further reinforces these patterns across activity spaces.
People's preferences, whether leaning towards traditional or achievement-oriented values, also shape their activity spaces. \par

\textbf{Physical and spatial factors}.
The design of urban spaces, including the layout of neighborhoods, proximity to amenities, and public transportation systems, influence experienced segregation.
Therefore, housing policies, urban planning decisions, and transport infrastructure can potentially mitigate or exacerbate spatial segregation. 
These elements determine where people can reside, work, and participate in daily activities, significantly influencing their experienced segregation. \par

\subsection{Future research directions}
Numerous studies apply activity space approaches to quantify socio-spatial segregation, utilizing both small and large data sets. 
These approaches are predominantly descriptive, focusing on defining and measuring socio-spatial segregation through human mobility data. 
Their principal contribution lies in transcending traditional residential perspectives. 
With over a decade since this paradigm shift in socio-spatial segregation research, moving from a static residential viewpoint to a dynamic, mobility-based one, the field is now poised to advance beyond mere descriptive analysis. 
There is a pressing need to explain observed social segregation phenomena within various spatiotemporal contexts using more causal, counterfactual, and hypothesis-testing methods. \par

As highlighted in this review (Section \ref{sec:fact_prom}), built environment studies have provided insights into segregation in diverse spatial contexts over time. 
However, these studies often treat socio-spatial segregation primarily as a static residential phenomenon, seldom considering how individuals go about their daily lives. 
For instance, in transport equity evaluations, accessibility is primarily calculated based on residential locations. 
Despite these efforts, the direct contributions of these factors to individuals' experienced segregation remain somewhat ambiguous. \par

Looking ahead, we advocate for three critical research directions emphasizing the need for a cross-disciplinary approach. 
These include exploring experienced segregation and devising region-specific explanations. 
First, we suggest that activity space approaches fueled by big geodata should be integrated with additional data sources that quantify transport systems and urban spaces. 
This integration would enable a more comprehensive analysis of the relationships among housing, transport access, urban design, and individual experienced segregation \citep{zhang2022residential,vachuska2023racial, nilforoshan2023human}, thereby maximizing the potential of big data's scale effect.
Second, studies employing urban space analysis to foster spatial integration should incorporate empirical insights into people's mobility behaviors.
As empirical mobility data reveals, the urban design challenge for promoting social inclusion may reside in bridging the gap between intended and observed co-presence between different population groups.
Third, one shall explore the causal relationships between land use, transportation infrastructure, and experienced segregation. 
Investigating how variations in land use patterns and accessibility levels influence human mobility could yield valuable insights into the effectiveness of policy interventions that reduce socio-spatial segregation. 
This approach could help identify strategies for enhancing community cohesion through urban planning and policy design.

\section*{Acknowledgements}
During the preparation of this work the authors used GPT-4 and Jenni.ai in order to improve the language use. 
After using this tool/service, the authors reviewed and edited the content as needed and take full responsibility for the content of the publication.

\appendix
\renewcommand{\thefigure}{A.\arabic{figure}}
\renewcommand{\thetable}{A.\arabic{table}}
\setcounter{figure}{0}
\setcounter{table}{0}
\section{Literature summary}\label{seca:method}
\begin{figure*}[!htp]
\centering
\includegraphics[width=1\textwidth]{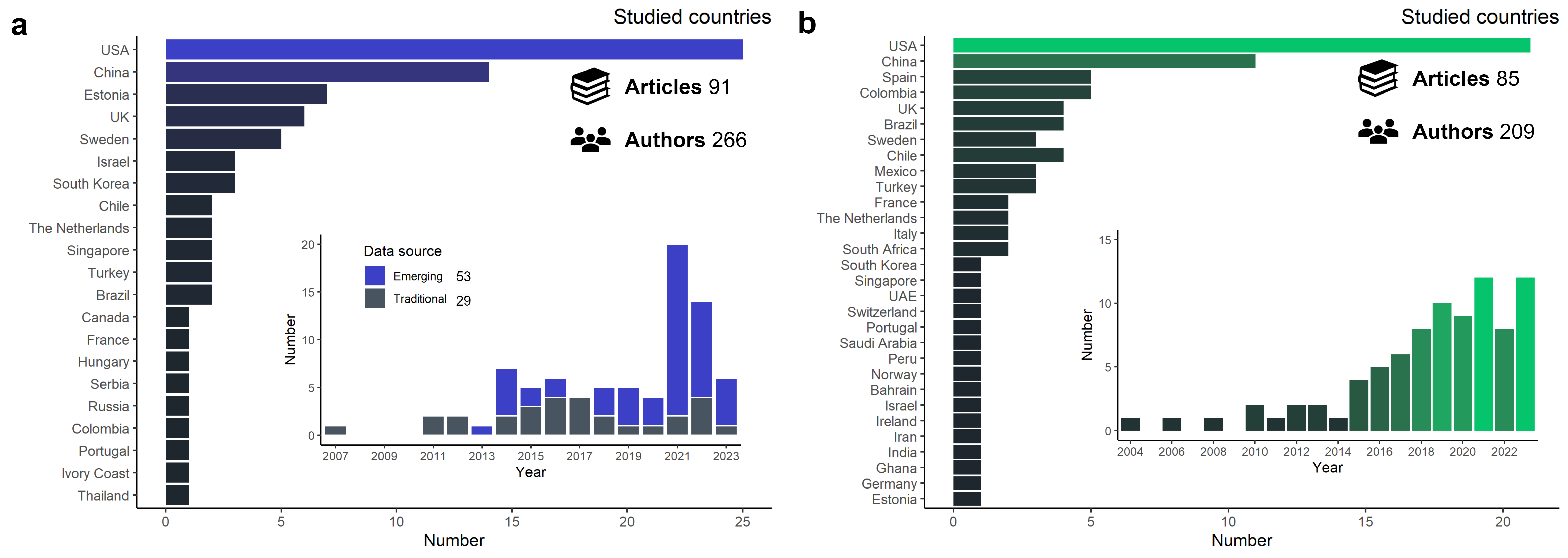}
  \caption{Referenced papers from activity space and built environment perspectives. These statistics only include original studies published in journals or conference proceedings, excluding reviews, opinions, and other articles cited in this study. (a) Number of activity space papers by studied country, year, and data source (Themes 1 \& 2.) (b) Number of transport and urban science papers by studied country and year (Themes 1 \& 3 and Themes 1 \& 4.)}\label{fig:refereces}
\end{figure*}

Figure \ref{fig:refereces} summarizes the included original studies.
Figures \ref{fig:refereces}a-b show the studies cited under Section \ref{sec:exp} on empirical findings of experienced segregation vs. residential segregation.
Figures \ref{fig:refereces}c-d illustrate those references under Section \ref{sec:discussion} on explaining experienced segregation from a built environment perspective. \par

We summarize the core concepts defined and used in this study in Table \ref{tab:concepts}. \par

\begin{table*}[!ht]
\caption{Glossary of core concepts.}\label{tab:concepts}
\centering
\begin{tabularx}{1\textwidth}{p{4cm}Xp{5.5cm}}
\hline
Concept  & Definition & Similar concepts \\\hline
Co-presence               & The sharing of place/urban space between individuals. Originated from space syntax theory \citep{hillier2007city}. & Exposure \citep{nilforoshan2023human}, Encounter \citep{rokem2018segregation}, Mixing \citep{juhasz2023amenity}, Co-existence \citep{csevik2022coexistence}, Co-location \citep{deurloo2022co}, Population exposed \\
Interaction (social)      & The process by which individuals act and react in relation to others within a social context.  & Contact/``Distanced'' interaction \citep{allport1954nature}  \\
Place/urban space         & A location within a city defined by its physical, social, or functional characteristics, e.g., parks.  & Amenity \citep{juhasz2023amenity}  \\
Areal unit & A defined area or segment of space used as the basis for segregation analysis, e.g., census tracts and administrative divisions.  & Spatial unit, Neighborhood \\
Mobility & The movement of individuals in space and time \citep{barbosa2018human}. & Human mobility \\
Activity space            & A geographic space encompassing an individual’s activity locations and movement over time  \citep{golledge1997spatial}.  & Daily path \citep{hagerstrand1975space} \\
Socio-spatial segregation & The spatial separation between population groups, suggesting their uneven spatial distributions and a lack of inter-group interactions \citep{yao2019spatial}. & Isolation, Discrimination \\
Social segregation & The extent of isolation and separation among population groups, limiting their interactions and social connections. & \\
Residential segregation   & The uneven spatial distribution of the residential location of different groups within a city or metropolitan region. & Housing segregation \\
Network segregation       & The uneven spatial distribution of the potential co-presence in locations different socio-economic groups can reach from home through walk, car, or transit networks. & Centrality measures of spatial network \citep{rokem2018segregation} \\
Place-based segregation   & The spatial separation of different groups within distinct geographic areas or locations. & Workplace segregation \citep{hellerstein2008workplace} \\
Visiting segregation      & The spatial separation of different groups within distinct geographic areas or locations, considering the time-varying co-presence. & Mobility-based segregation \citep{park2021we,iyer2024mobility} \\
Experienced segregation   & How segregated an individual is across his/her activity space. & Multi-contextual segregation \citep{park2018beyond}  \\
Lifestyle & The way a person or a group chooses to live. & Habits, Behavior \\
Homophily & Individuals are more likely to be co-present with others from similar backgrounds \citep{amini2014impact,xu2022beyond}. & Homophilic mobility \\\hline
\end{tabularx}
\end{table*}

\renewcommand{\thefigure}{B.\arabic{figure}}
\renewcommand{\thetable}{B.\arabic{table}}
\setcounter{figure}{0}
\setcounter{table}{0}
\section{Segregation metrics and implications from practices}\label{secb:measure_metr}
A variety of metrics are used to quantify socio-spatial segregation.
In Table \ref{tab:metrics}, we present a selection of segregation metrics, categorized into three types based on their adoption in the three approaches defined in this study.
The first class (Type 1) consists of traditional segregation metrics developed for residential segregation, which can also be used for evaluating the co-presence outside homes in activity space approaches. 
The second class (Type 2) considers network centrality metrics to quantify potential co-presence, e.g., the likelihood of different population groups occupying or traversing the same spaces within a spatial network \citep{legeby2015streets}. 
The third class (Type 3) comprises metrics commonly seen in studies using activity space approaches.
\par

\begin{table*}[!ht]
\small
\caption{Selection of key segregation metrics. Type I = Classic, developed for quantifying residential segregation. Type II = Network analysis, applied in built environment research in approximating the potential co-presence of individuals. Type III = Activity space, applied or developed in the context of activity space approaches, often along with a large amount of geolocation data on human mobility. }\label{tab:metrics}
\centering
\begin{tabularx}{1\textwidth}{llX}
\hline
Type                & Index/Model                                  & Description                                                                                                            \\\hline
\multirow{5}{*}{I}  & Evenness or Dissimilarity                    & How evenly groups are distributed across a geographical space \citep{reardon2004measures}.                                    \\
                    & Exposure Indices                             & Potential contact between groups \citep{silm2014temporal}.                                                                      \\
                    & Concentration Indices                        & The extent to which minority populations are situated in specific regions \citep{johnston2007ethnic}.                         \\
                    & Spatial Distribution Indices                 & Broader spatial patterns \citep{demoraes2021live}.                                          \\
                    & Spatial Clustering Indices                   & The degree of clustering of high-density group areas, e.g., Moran's I \citep{o2007surface}.                                  \\\hline
\multirow{7}{*}{II}  & Integration Analysis/Closeness                & Connectedness of each space highlighting areas likely to be frequented by diverse groups \citep{yunitsyna2023investigating}.  \\
                    & Choice Analysis/Betweenness                  & It quantifies a space's role in facilitating encounters \citep{rokem2018segregation}.                                           \\
                    & Visibility Graph Analysis                    & Which spaces are visually connected, providing insight into interaction opportunities \citep{turner2001isovists}.             \\
                    & Angular Segment Analysis                     & Spaces accessible with fewer angular changes often being busier and thus prime for co-presence \citep{turner2001angular}.         \\
                    & Combined integration and choice values       & They identify the network's most accessible areas \citep{rokem2018segregation}.                                                 \\
                    & Multi-accessibility                          & Simultaneous ease of access to a place by multiple transport modes \citep{carpio2021multimodal}.   \\\hline
\multirow{10}{*}{III} & Socio-Spatial Isolation Indices              & Individual isolation within activity locations \citep{athey2021estimating}.                                                        \\
                    & Social Interaction Potential                 & Metric of exposure based on time-geography for metropolitan scale \citep{farber2015measuring}.                            \\
                    & i-STP index                                  & Individual-level segregation index considering different times of the day \citep{park2018beyond}.                             \\
                    & Flow-based spatial interaction model         & It can capture the impact of specified commuting routes on segregation experiences \citep{shen2019segregation}.               \\
                    & Segregation hotspots                         & Cluster multiscalar fingerprint to identify hotspots of segregation in urban spaces \citep{olteanu2020clustering}.            \\
                    & Segregated Mobility Index (SMI)              & Racial segregation in how neighborhoods of varying racial compositions are connected \citep{candipan2021residence}.           \\
                    & Income Unevenness                            & Quantifies income unevenness in US cities using population income quartiles \citep{moro2021mobility}.        \\
                    & Graph Embedding                              & Income segregation that combines residential and mobility perspectives \citep{zhang2021discovering}.                            \\
                    & Index of Concentration at the Extremes (ICE) & How population concentrates at certain groups in a given area \citep{iyer2023mobility}.                                         \\
                    & Spatial Segregation Index                    & This index integrates distance-decay functions to measure individual experienced segregation \citep{wu2023revealing}.         \\\hline
\end{tabularx}
\end{table*}

In Table \ref{tab:actions}, we present a brief summary of policies and actions in housing, public transport, and urban design that have implications for segregation.

\begin{table*}[!ht]
\caption{Policies and actions in housing, public transport, and urban design and their impact on segregation. 1 = Housing, 2 = Public transport, and 3 = Urban design.}\label{tab:actions}
\centering
\begin{tabularx}{1\textwidth}{p{0.5cm}p{4cm}p{5cm}X}
\hline
\multicolumn{1}{l}{\#}        & Policy/Action                                             & Description                                                & Segregation implications                                       \\\hline
\multirow{3}{*}{1}          & Urban village placement for rural migrant workers (China) & Good job accessibility for rural workers in urban villages. & Lowers workplace segregation \citep{zhu2017residential,zhou2021workplace}. \\
                                  & Urban renewal (China)                                     & Urban village demolition due to renewal policies.           & Risks increasing segregation due to displacement \citep{zhu2022residential}.\\
                                  & Housing Choice Voucher program (US)                       & Housing affordability programs.                             & Effectiveness limited by entrenched socio-demographic barriers \citep{garboden2021you}.\\\hline
\multirow{4}{*}{2} & Minibus taxi system (South Africa)                        & Integrated transport systems for diverse populations.       & Could decrease segregation by serving actual transport needs \citep{nelson2023spatial}.\\
                                  & Public transport fare reform (Portugal)                   & Enhancing commuter accessibility.                           & May reduce social inequalities and segregation \citep{silver2023inequality}.\\
                                  & Bus and metro systems (Colombia)                          & Public transport accessible to the poor.                    & Reduces segregation by improving transport access for the poor \citep{valenzuela2023income}.\\
                                  & Light-rail transportation (US)                            & Light-rail construction in neighborhoods.                   & Gentrification risk with potential for demographic shifts \citep{hess2020light}.\\\hline
\multirow{5}{*}{3}     & Bike share program (US)                                   & Equitable bike station planning.                            & Equitable transport access can lower segregation \citep{bhuyan2019gis}.\\
                                  & Cycle hire scheme (UK)                                    & Wider distribution of city cycling schemes.                 & Enhanced inclusiveness and reduced income segregation \citep{lovelace2020london}.\\
                                  & Inclusive sidewalk design (Saudi Arabia)                  & Gender-responsive public space design.                      & Targets gender-based segregation reduction \citep{almahmood2018sidewalk}.\\
                                  & Urban transformation (Turkey)          & Social inclusion efforts for refugees.                      & Fosters integration and combats segregation in mixed-use areas \citep{altaema2023urban}.\\
                                  & Carpooling (US)                                           & Carpooling algorithms for diverse groups.                   & Encourages social integration   via shared transport \citep{librino2020home}. \\\hline
\end{tabularx}
\end{table*}

\bibliographystyle{unsrtnat}
\bibliography{references}

\end{document}